\documentstyle[12pt]{article}
\begin{document}
\title{Higher Derivative Gravitation in Superstrings}
\author{Burt A. Ovrut\thanks{Talk presented at STRINGS'95, University of
Southern California, Los Angeles, CA, March 13-18, 1995. Work supported in
part by DOE Contract DOE-AC02-76-ERO-3071 and NATO Grant CRG. 940784.}
\\Department of Physics, University of Pennsylvania\\
Philadelphia, PA 19104-6396, USA}
\date{}
\maketitle
\begin{abstract}
A discussion of the number of degrees of freedom, and their dynamical
properties, in higher derivative gravitational theories is presented. It is
shown that non-vanishing $(C_{mnpq})^{2}$ terms arise in N=1, D=4 superstring
Lagrangians due to one-loop radiative corrections with light field internal
lines.
\end{abstract}
\section{Bosonic Gravitation}
The usual Einstein theory of gravitation involves a symmetric tensor
$g_{\mu\nu}$ whose dynamics is determined by the Lagrangian
\begin{equation}
{\cal{L}}=-\frac{1}{2\kappa^{2}}{\cal{R}}
\label{eq:first}
\end{equation}
The diffeomorphic gauge group reduces the number of degrees of freedom from
ten down to six. Einsteins equations futher reduce the degrees of freedom to
two, which correspond to a physical spin-2 massless graviton. Now let us
consider an extension of Einstein's theory by including terms in the action
which are quadratic in the curvature tensors. This extended Lagrangian is given
by
\begin{equation}
{\cal{L}}=
-\frac{1}{2\kappa^{2}}{\cal{R}}+\alpha{\cal{R}}^{2}+\beta(C_{mnpq})^{2}
+\gamma({\cal{R}}_{mn})^{2}
\label{eq:second}
\end{equation}
${\cal{R}}^{2}$, $(C_{mnpq})^{2}$, and $({\cal{R}}_{mn})^{2}$ are a complete
set
of CP-even quadratic curvature terms. The topological Gauss-Bonnet term is
given by
\begin{equation}
GB=(C_{mnpq})^{2}-2({\cal{R}}_{mn})^{2}+\frac{2}{3}{\cal{R}}^{2}
\label{eq:third}
\end{equation}
Therefore, we can write
\begin{equation}
{\cal{L}}=-\frac{1}{2\kappa^{2}}{\cal{R}}+a{\cal{R}}^{2}+b(C_{mnpq})^{2}+cGB
\label{eq:fourth}
\end{equation}
In this case, it can be shown \cite{A} that there is still a physical spin-2
massless
graviton in the spectrum. However, the addition of the ${\cal{R}}^{2}$ term
introduces  a new physical spin-0 scalar, $\phi$, with mass
$M=\frac{1}{\sqrt{a}\kappa}$. Similarly, the $(C_{mnpq})^{2}$
term introduces a spin-2 symmetric tensor, $\phi_{mn}$, with mass
$M=\frac{1}{\sqrt{b}\kappa}$ but this field, having wrong sign
kinetic energy, is ghost-like. The GB term, being a total divergence, is
purely topological and does not lead to any new degrees of freedom. The
scalar $\phi$ is perfectly physical and can lead to very interesting new
physics \cite{B}. The new tensor $\phi_{mn}$, however, appears to be
problematical.
There have been a number of attempts to show that the ghost-like behavior of
$\phi_{mn}$ is allusory, being an artifact of linearization \cite{C}. Other
authors
have pointed out that since the mass of $\phi_{mn}$ is near the Planck scale,
other Planck scale physics may come in to correct the situation \cite{D}. In
all these attempts, the gravitational theories being discussed were not
necessarily consistent and well defined. However, in recent years, superstring
theories have emerged as finite, unitary theories of gravitation.
Superstrings, therefore, are an ideal laboratory for exploring the issue of
the ghost-like behavior of $\phi_{mn}$, as well as for asking whether the
scalar
$\phi$ occurs in the superstring Lagrangian. Hence, we want to explore the
question ``Do Quadratic Gravitation Terms Appear in the N=1, D=4 Superstring
Lagrangian''?
\section{Superspace Formalism}
In the Kahler (Einstein frame) superspace formalism, the most general
Lagrangian for Einstein gravity, matter and gauge fields is
\begin{equation}
{\cal{L}}_{E}
=-\frac{3}{2\kappa^{2}}\int{d^{4}{\theta}E[K]}+\frac{1}{8}\int{d^{4}
\theta\frac{E}{R}f(\Phi_{i})_{ab}W^{{\alpha}a}W_{\alpha}^{b}} + hc
\label{eq:fifth}
\end{equation}
where we have ignored the superpotential term which is irrelevant for this
discussion. The fundamental supergravity superfields are R and
$W_{\alpha\beta\gamma}$, which are chiral, and $G_{\alpha{\dot\alpha}}$, which
is
Hermitian. The bosonic ${\cal{R}}^{2}$, $(C_{mnpq})^{2}$ and
$({\cal{R}}_{mn})^{2}$ terms are contained in the highest components of the
superfields $\bar{R}R$, $(W_{\alpha\beta\gamma})^{2}$ and
$(G_{\alpha\dot{\alpha}})^{2}$ respectively. One can also define the
superGauss-Bonnet combination
\begin{equation}
SGB=8(W_{\alpha\beta\gamma})^{2}+(\bar{\cal{D}}^{2}-8R)(G_{\alpha\dot{\alpha}}
^{2}-4\bar{R}R)
\label{eq:sixth}
\end{equation}
The bosonic Gauss-Bonnet term is contained in the highest chiral component of
SGB. It follows that the most general quadratic supergravity Lagrangian is
given by
\begin{eqnarray}
{\cal{L}}_{Q}&=&\int{d^{4}\theta E
\Sigma(\bar{\Phi}_{i},\Phi_{i})\bar{R}R}+\int
{d^{4}\theta\frac{E}{R}g(\Phi_{i})(W_{\alpha\beta\gamma})^{2}} \\ \nonumber
{ }&+&\int{d^{4}\theta
E\Delta(\bar{\Phi}_{i},\Phi_{i})(G_{\alpha\dot{\alpha}})^{2}}+ hc
\label{eq:seventh}
\end{eqnarray}
\section{(2,2) Symmetric $Z_{N}$ Orbifolds}
Although our discussion is perfectly general, we will limit ourselves to
orbifolds, such as $Z_{4}$, which have (1,1) moduli only. The relevant
superfields are the dilaton, S, the diagonal moduli $T^{II}$, which we'll
denote as $T^{I}$, and all other moduli and matter superfields, which we
denote collectively as $\phi^{i}$. The associated Kahler potential is
\begin{displaymath}
K=K_{0}+Z_{ij}\bar{\phi}^{i}\phi^{j}+{\cal{O}}((\bar{\phi}\phi)^{2})
\end{displaymath}
\begin{equation}
\kappa^{2}K_{0}=-ln(S+\bar{S})-\sum{(T^{I}+\bar{T}^{I})}
\label{eq:eigth}
\end{equation}
\begin{displaymath}
Z_{ij}=\delta_{ij}\prod{(T^{I}+\bar{T}^{I})^{q^{i}_I}}
\end{displaymath}
The tree level coupling functions $f_{ab}$ and g can be computed uniquely from
amplitude computations and are given by
\begin{equation}
f_{ab}=\delta_{ab}k_{a}S
\label{eq:ninth}
\end{equation}
\begin{displaymath}
g=S
\end{displaymath}
There is some ambiguity in the values of $\Delta$ and $\Sigma$ due to the
ambiguity in the definition of the linear supermultiplet. We will take the
conventional choice
\begin{equation}
\Delta=-S
\label{eq:tenth}
\end{equation}
\begin{displaymath}
\Sigma=4S
\end{displaymath}
It follows that, at tree level, the complete $Z_{N}$ orbifold Lagrangian is
given by ${\cal{L}}={\cal{L}}_{E}+{\cal{L}}_{Q}$ where
\begin{equation}
{\cal{L}}_{Q}=\frac{1}{4}\int{d^{4}\theta\frac{E}{R}SSGB}+ hc
\label{eq:eleventh}
\end{equation}
Using this Lagrangian, we now compute the one-loop moduli-gravity-gravity
anomalous threshold correction \cite{E}. This must actually be carried out in
the
conventional (string frame) superspace formalism and then transformed to
Kahler superspace \cite{F}. We also compute the relevant superGreen-Schwarz
graphs.
Here we will simply present the result. We find that
\begin{displaymath}
{\cal{L}}_{massless}^{1-loop}
=\frac{1}{24(4\pi)^{2}}\Sigma\left[h^{I}\int d^{4}\theta
(\bar{\cal{D}}^{2}-8R)\bar{R}R\frac{1}{\partial^{2}}D^{2}ln(T^{I}+\bar{T}^{I})
\right.
\end{displaymath}
\begin{equation}
+(b^{I}-8p^{I})
\int{d^{4}\theta(W_{\alpha\beta\gamma})^{2}\frac{1}{\partial^{2}}
D^{2}ln(T^{I}+\bar{T}^{I})}
\label{eq:twelveth}
\end{equation}
\begin{displaymath}
\left.+p^{I}\int{d^{4}\theta(8(W_{\alpha\beta\gamma})^{2}+
(\bar{\cal{D}}^{2}-8R)
((G_{\alpha\dot{\alpha}})^{2}-4\bar{R}R))\frac{1}{\partial^{2}}
D^{2}ln(T^{I}+\bar{T}^{I})}+hc\right]
\end{displaymath}
where
\begin{displaymath}
h^{I}=\frac{1}{12}(3\gamma_{T}+3\vartheta_{T}q^{I}+\varphi)
\end{displaymath}
\begin{equation}
b^{I}=21+1+n_{M}^{I}-dimG+\Sigma(1+2q_{I}^{i})-24\delta_{GS}^{I}
\label{eq:thirteenth}
\end{equation}
\begin{displaymath}
p^{I}=-\frac{3}{8}dimG-\frac{1}{8}-\frac{1}{24}\Sigma1+\xi-3\delta_{GS}^{I}
\end{displaymath}
The coefficients $\gamma_{T}$ and $\vartheta_{T}$, which arise from moduli
loops,
and $\varphi$ and $\xi$, which arise from gravity and dilaton loops, are
unknown.
However, as we shall see, it is not necessary to know their values to
accomplish our goal. Now note that if $h^{I}\neq0$ then there are
non-vanishing ${\cal{R}}^{2}$ terms in the superstring Lagrangian. If
$b^{I}-8p^{I}\neq0$ then the Lagrangian has $C^{2}$ terms. Coefficient
$p^{I}\neq0$ merely produces a Gauss-Bonnet term. With four unknown parameters
what can we learn? The answer is,a great deal! Let us take the specific
example of the $Z_{4}$ orbifold. In this case, the Green-Schwarz coefficients
are known cite{G}
\begin{equation}
\delta_{GS}^{1,2}=-30,  \delta_{GS}^{3}=0
\label{eq:fourteen}
\end{equation}
which gives the result
\begin{equation}
b^{1,2}=0,  b^{3}=11\times24
\label{eq:fifteen}
\end{equation}
Now, let us try to set the coefficients of the $(C_{mnpq})^{2}$ terms to zero
simultaneously. This implies that
\begin{equation}
b^{I}=8p^{I}
\label{eq:sixteen}
\end{equation}
for I=1,2,3 and therefore that
\begin{equation}
p^{1,2}=0,  p^{3}=33
\label{eq:seventeen}
\end{equation}
{}From this one obtains two separate equations for the parameter $\xi$ given by
\begin{equation}
\xi=\frac{3}{8}dimG+\frac{1}{8}+\frac{1}{24}\Sigma1-90
\label{eq:eigthteen}
\end{equation}
for I=1,2 and
\begin{equation}
\xi=\frac{3}{8}dimG+\frac{1}{8}+\frac{1}{24}\Sigma1
\label{eq:nineteen}
\end{equation}
for I=3. Clearly these two equations are incompatible and, hence, it is
impossible to have all vanishing $(C_{mnpq})^{2}$ terms in the 1-loop
corrected Lagrangian of $Z_{4}$ orbifolds. We find that the same results hold
in other orbifolds as well.
\section{Conclusion}
We conclude that non-vanishing $(C_{mnpq})^{2}$ terms are
generated by light field loops in the N=1, D=4 Lagrangian of $Z_{N}$ orbifold
superstrings. It is conceivable that loops containing the heavy tower of
states might cancel these terms, but we find no reason, be it duality or any
other symmetry, for this to be the case \cite{H}. This is presently being
checked by a
complete genus-one string calculation \cite{I}. We conjecture that cancellation
will
not occur. Unfortunately, since $\gamma_{T}$,
$\vartheta_{T}$ and $\varphi$ are unknown, we can say nothing concrete about
the
existence of ${\cal{R}}^ {2}$ terms in the Lagrangian. This issue will also be
finally resolved in the complete superstring calculation.

\end{document}